\documentclass[fleqn,10pt]{SelfArx} 

\usepackage{lipsum} 

\definecolor{color1}{RGB}{0,0,90} 
\definecolor{color2}{RGB}{0,20,20} 

\usepackage{hyperref} 
\hypersetup{hidelinks,colorlinks,breaklinks=true,urlcolor=color2,citecolor=color1,linkcolor=color1,bookmarksopen=false,pdftitle={Title},pdfauthor={Author}}

\JournalInfo{Accepted to Astronomy Reports, 2018, Vol. 62, No. 8} 
\Archive{}

\PaperTitle{
Superflares on Giant Stars
} 

\Authors{M.\,M.~Katsova\textsuperscript{1}*, L.\,L.~Kitchatinov\textsuperscript{2,3}, D.~Moss\textsuperscript{4},
K.~Ol\'ah\textsuperscript{5}, D.\,D.~Sokoloff\textsuperscript{6,7}
} 

\affiliation{\textsuperscript{1}\textit{
Sternberg State Astronomical Institute, Lomonosov Moscow State University, 119991, Moscow, Russia}} 
\affiliation{\textsuperscript{2}\textit{
Institute for Solar-Terrestrial Physics, PO Box 291, Irkutsk, 664033, Russia
}} 
\affiliation{\textsuperscript{3}\textit{
Pulkovo Astronomical Observatory, St. Petersburg 196140, Russia
}} 
\affiliation{\textsuperscript{4}\textit{
School of Mathematics, University of Manchester, Manchester M13 9PL, UK
}} 
\affiliation{\textsuperscript{5}\textit{
Konkoly Observatory, Research Centre for Astronomy and Earth Sciences, Hungarian Academy of Sciences, 1121 Budapest, 
Konkoly Thege M. u., 15-17, Hungary
}} 
\affiliation{\textsuperscript{6}\textit{
Department of Physics, Moscow State University, Moscow, 119991, Russia
}} 
\affiliation{\textsuperscript{7}\textit{
IZMIRAN, 4 Kaluzhskoe Shosse, Troitsk, Moscow, 142190, Russia
}} 
\affiliation{*\textbf{Corresponding author}: maria@sai.msu.ru} 

\newcommand{\keywordname}{Keywords} 

\Abstract{
The Kepler mission identified huge flares on various stars including some of solar type.
These events are substantially more energetic than solar flares, and  so they are referred to as 
superflares.
Even a small probability of such a superflare on the Sun would be a menace to modern society.
A  flare comparable in energy with that of superflares was observed on 24th and 25th September on the binary 
HK Lac.
Unlike the Kepler stars, there are observations of differential rotation for HK Lac. This differential 
rotation appears to be anti-solar. For anti-solar differential rotation, dynamo models can give magnetic 
activity waves of dipole symme\-try
as well as quasi-stationary magnetic configurations with quadrupole symme\-try. The magnetic energy of such 
stationary configurations is usually about two orders of magnitude higher than that associated with 
activity waves. We believe that this mechanism could provide sufficient energy to produce superflares on 
late type stars, and present some simple models in support of this idea. 
}

\begin{document}

\flushbottom 

\maketitle 


\thispagestyle{empty} 


\section*{Introduction} 

\addcontentsline{toc}{section}{Introduction} 

Data from the {\sl Kepler} mission \cite{Kea10} has revealed the existence  of stellar 
"superfla\-res" 
\cite{Mea12,Sea13}.
Some of the superflaring stars are  G dwarfs
with long rotation periods $P_{\rm rot} > 10$ days \cite{Sea13}, and some seem to be very similar 
to the Sun in their surface temperature
and rotation rate, \cite{Nea14}. 
The energies of stellar flares observed belong to in the wide range
from $10^{33}$ to $10^{37}\;$erg  \cite{Sea13}\  while the highest energy of any  observed solar flares is
approximately $(2 - 3) \times 10^{32}\;$erg.

Observations of superflares on Sun-like stars are challenging for solar and stellar physics in general and 
for the stellar dynamo in particular. 
A dynamo scenario to explain the accumulation of magnetic energy sufficient to 
support super\-flares has been recently suggested by \cite{KO16} and is further developed by 
\cite{Kea18}. The idea of this scenario is that  dynamo action 
on superflaring stars, instead of producing stellar 
cycles  similar to the solar Schwabe 11-year cycle, excites a quasi-stationary magnetic configuration with  
a much higher magnetic energy, simply 
because the dynamo drivers only have to build and maintain a steady
field, whereas the solar dynamo has also to drive magnetic field reversals 
during  the course of the cycle.  The 
scenario proposed by \cite{KO16} suggests that a quasi-stationary behaviour appears for a limited 
time as a result of fluctuations in the dynamo governing parameters while \cite{Kea18} considers that 
quasi-stationary behaviour is a normal magnetic field configuration in superflaring stars. Observa\-tions of 
the {\sl Kepler} superflaring stars are consistent with these interpretations.
Certainly they  do not provide sufficient information for 
its verification, nor do they allow choice between these scenarios.

Remarkably, there is a giant, HK Lac, for which a very strong optical event is known from ground-based optical 
observations, and observational information required for understanding of the 
origin of the strong magnetic field origin.

Observing superflares for magnetoactive evolved giant stars is challenging as well.  From the very high 
precision {\sl Kepler} data \cite{VSD17} found 653 flaring giant stars from the Kepler sample of 22837 
giants, which means 2.86\% flare incidence. Comparing to the flare incidence of 2.90\% and 5.28\% for the G 
and K--M dwarfs respectively,  there is little difference. 
For giants generally, the flare energies are somewhat higher and the durations are longer than those 
observed on dwarfs \cite{VSD17}.

From the ground however, the observational  scatter 
is much larger than that of the {\sl Kepler} observations, and so the flares are usually 
outshone by the brightness of the giant stars, i.e. the low amplitude events simply disappear in the 
observational scatter, so only the rare, most energetic ones, can be obtained. Moreover, the ground 
based observations do not provide continuous monitoring, rather one or a few 
datapoints per night are observed. Summing 
up, the probability of observing a superflare on a giant star from the ground is very low, but fortunately 
there are nevertheless a (very) few examples.

The aim of our paper is to attract attention in this context to the 
strong $H_\alpha$ flare that occurred on HK Lac on the 24th and 25th September,
1989 \cite{CF94}. In our opinion, the observational data 
concerning this flare support the {\sl Kepler} results for some 
giant superflaring stars \cite{B15,VSD17}, 
as well as the corresponding dynamo interpretation suggested by \cite{Kea18}.

Superflares  occur
on close binaries like HK Lac. A higher flaring frequency at epochs of periastron passage in this type of 
star \cite{MMN02} implies their binary nature to be somehow related to superflares. Recent analysis by 
\cite{L17} however shows that binarity can provoke superflares if sufficiently high magnetic anergy is 
available but it is not the reason for high magnetic energy. This is why we seek dynamo mechanisms producing 
the abnormally high magnetic energy.

In order to facilitate this comparison we present the data concerning HK Lac 
as a superflaring star following the logic of the {\sl Kepler} mission, 
i.e. from the data relating a particular superflare to the data concerning physical parameters of the star 
and then to the dynamo model.

On one hand, the origin of superflares on giant stars is interesting in itself. On the other hand 
however, the data concerning superflaring giant stars can help us to a 
better understanding of the origin of 
superflares on solar-like stars. The point is that we can here use observational data concerning the 
differential rotation of HK Lac. In this respect, the case of HK Lac plays  an exceptional role in the 
following.

\section{Basic parameters and differential rotation of HK Lac}

HK Lac is the K0 III primary star  ($T_{\rm eff} \approx 4700\;$K) of an RS CVn-type binary system with an unknown 
secondary star. The early history of the binary since its discovery until the observed $H_\alpha$ flares is 
summarized by \cite{Oea97} and references therein. The mass of HK Lac is not known; for more 
information on the problem of deriving masses of highly active giants -- see e.g. \cite{Oea14}.

Briefly, HK Lac is the primary star of a close binary system with an orbital period of 24.4284 days and mass 
function of 0.105  \cite{GH71}.  The orbital eccentricity is close to zero so the system 
is tidally locked,
therefore the observed rotational period is close to the orbital. The rotational period  
is changing slightly because of the differential rotation. Using Doppler imaging techniques the differential 
rotation of HK Lac is found to be of anti-solar type with $\alpha = -0.05 \pm 0.05$ \cite{K17} and \cite{Wetal05}, see their Fig. 2, though the 
value is uncertain. The very low {\it absolute} value of the differential rotation of HK Lac is probably due 
to its binary nature similar to that of others on giant RS CVn binaries in close, tidally locked systems \cite{K17}.

Strong chromospheric activity was observed on HK Lac by the IUE satellite between about 1300--3000\;\AA , 
including the MgII  h\&k lines. Strong turbulent motions in the chromosphere were indicated by the widths of 
these lines above the rotational broadening \cite{Oea92}. HK Lac has multiple and changing spot 
cycles between 5.4 and 5.9 years and 10 to 13.3 years \cite{Oea09}.  We believe that these cycles could 
be similar to the (also changing) solar cycles 
but however
do not require oscillating 
eigensolutions of dynamo equations responsible for the 11-year cycle.
This is a challenge for dynamo theory and also perhaps for
 observations and their interpretation.

Single giants of the same spectral type rotate much more slowly than the close binary primary of HK Lac \cite{G89}; nevertheless they can have large spots 
and show magnetic activity. This probably implies that in HK Lac the dynamo is highly supercritical.

\section{The superflare of 24th and 25th September 1989 on HK Lac}

According to \cite{CF94}, the total energy released in the $H_\alpha$ line during the event is estimated as  
$1.3 \times 10^{37}\;$erg, see also \cite{CF93}, and the estimated total released energy is in the order of 
$10^{39}\;$erg. This conclusion is based on the analysis of Fig.\ 2 from \cite{O91}. The superflare was 
localized at about the same longitude as the newly emerged large spotted area found by \cite{O91}. 
Similar coincidences of $H_\alpha$ flares and newly emerged spots were 
found in 1977 by \cite{CF94} -- Fig.\ 5 of that paper. 
The $H_\alpha$ observations by \cite{BT80} reported extreme emission variations in the $H_\alpha$ profile on 
Sept.\ 5, 1976, which were 
still present on the following day with lower flux. 
The remainder of their $H_\alpha$ data observed between 1976 and 1978 did not show such any strong emission.

Apart from the 6 day long $H_\alpha$ flare in 1989, two  other smaller flare-like events
in $H_\alpha$ were observed on HK Lac in 1993 and 1994, 
the latter lasting for  2 or 3 days, Fig. 2 in \cite{Cea96}. 

Sudden changes in surface activity suggest that giant long-lasting flares may have appeared during such abrupt changes.

\section{The dynamo modelling}

We believe that superflares as well as usual stellar and solar flares are associated with strong magnetic 
fields. Stellar magnetic fields are believed to be created by a dynamo. We have to show how this mechanism 
can create the huge magnetic energy supply required to produce such strong nonstationary processes and why this 
mechanism does not produce it on other stars, and on the Sun in particular.

Consider a 
dynamo driven in a spherical shell by differential rotation and mirror-asymmetric convection. The problem is 
that we cannot expect to know the internal 
hydrodynamics of this particular star with an accuracy comparable with that 
known for the Sun, especially since our available knowledge 
of internal solar hydrodynamics is still insufficient to reconstruct all 
the details of the solar dynamo. This 
is why we consider two possible examples of the rotation 
curve (Fig.~1), which are extrapolations of synthetic solar rotation laws
to a deep convective shell of 60\% of the stellar radius.

Red giants are known to possess extended convective envelopes. Asteroseis\-mology detects strong radial 
differential rotation with rotation rate increasing with depth in RGB stars \cite{Bea12,Tea17}. The 
subrotation can be related to predominant\-ly downward convective transport of angular momentum in slowly 
rotating single giants, cf. Fig.3. in \cite{K13}. Latitudinal transport can be more efficient in faster 
rotating (large Coriols/small Rossby numbers) binaries thus producing latitudinal rotational 
inhomogeneities.

Dynamos driven by joint action of differential rotation and mirror-asymmet\-ric convection are described by 
the following mean-field equations

\begin{equation}
\frac{\partial {\bf B}}{\partial t} = {\rm rot} \, (\alpha {\bf B} + {\bf V} \times {\bf B}) - \beta 
\; {\rm rot} \; {\rm rot} \; {\bf B} \, ,
\label{eq}
\end{equation}
where $\bf B$ is large-scale magnetic field, $\alpha$ is a measure for the mirror asymmetry of convection (the $\alpha$-effect), $\beta$ is convective diffusivity and $\bf V$ is large-scale velocity, here just differential rotation. Nonlinear dynamo saturation is described in the framework of the simplest model by algebraic $\alpha$-quenching.

The intensity of dynamo action in the model is controlled by a single paramet\-er, known as  the dynamo number 
$D$, which can be positive or negative. 
This 
sign distinguishes between solar and anti-solar types of differential rotation. 
The sign of $D$ depends on its definition and we use a definition which gives 
solar type differential rotation when $D<0$,  and anti-solar when $D>0$. 
In the course of the modelling we just change the sign of 
$D$ to move from solar to anti-solar differential rotation and keep the shape of the rotation law unchanged.

\begin{figure*} 
\begin{center} 
\large{(a)}
\includegraphics[width=0.75\linewidth]{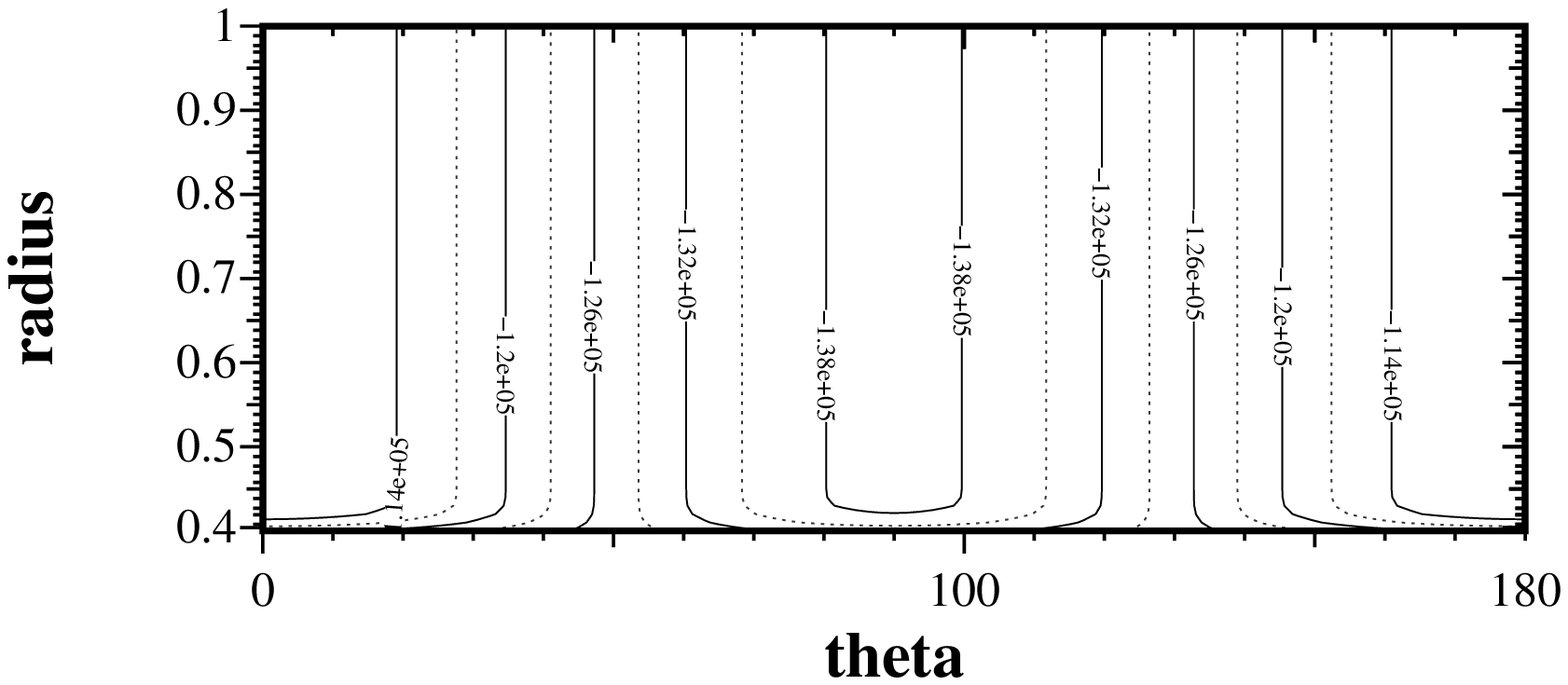}\\[0.3cm] 
\large{(b)}
\includegraphics[width=0.75\linewidth]{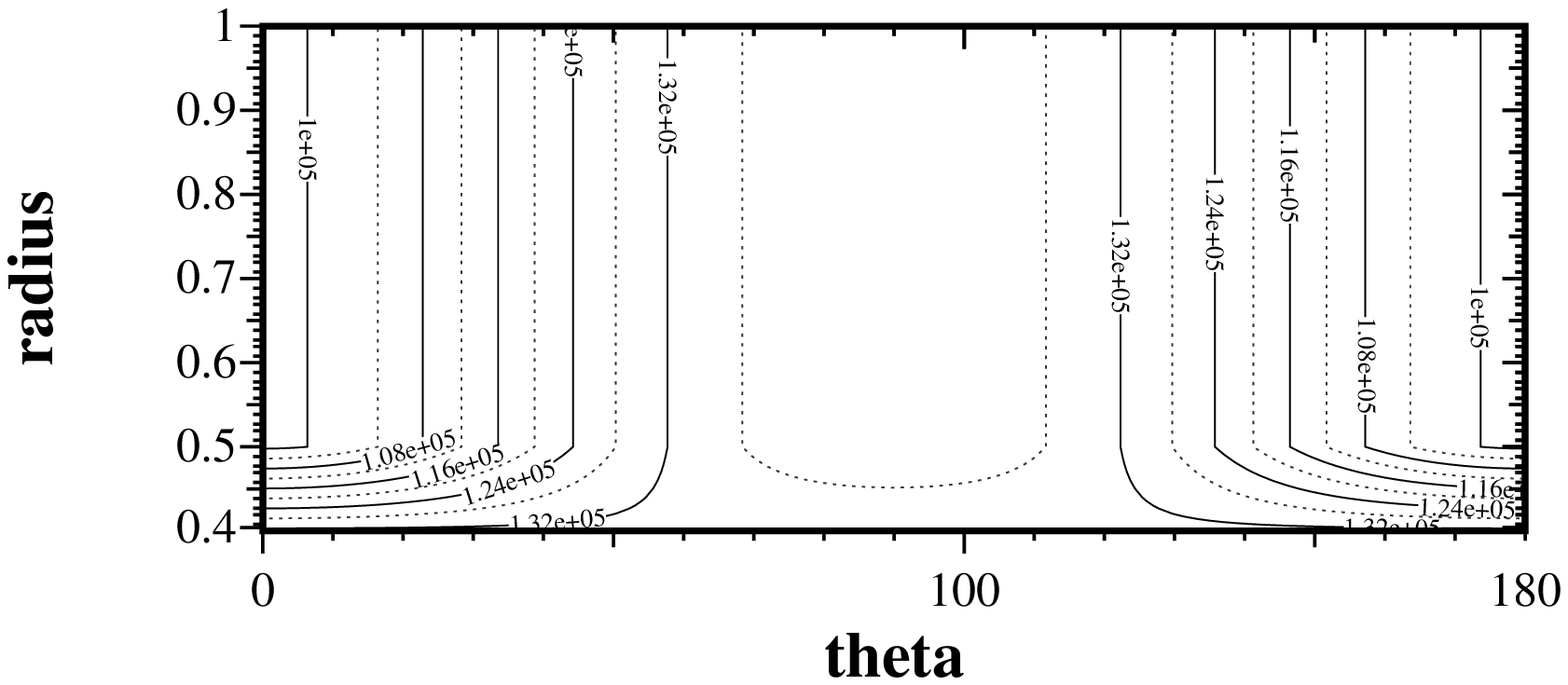}\\[0.5cm] 
\caption{Rotation laws used in the dynamo modelling. a) modified  \cite{Jea08} rotation curve with the nominal
width of the convective shell 0.60 stellar radius; b) modification of 
an alternative synthetic solar rotation law.}
\label{rotcurve}
\end{center}
\end{figure*}
 
According to  \cite{G89PASP}, magnetic activity occurs in typical giant stars which rotate 
much more slowly than HK Lac. If in the former stars $D$ is sufficiently 
supercritical to support dynamo action, it is natural to expect that the
 dynamo number for HK Lac is 
highly supercritical, possibly 10 times larger than the critical value. \cite{KO11}  explain how a highly supercritical value of  $D$  might be 
compatible with smaller fractional differential  stellar rotation  (in fact the  ratio of toroidal 
field winding time to diffusion time  is more important here; our simple models do not take account of this).  E.g., red dwarf stars 
smaller than the Sun also have smaller differential rotation but are also more dynamo-efficient \cite{KO11}.

\begin{figure*} 
\centering
a)\includegraphics[width=0.45\linewidth]{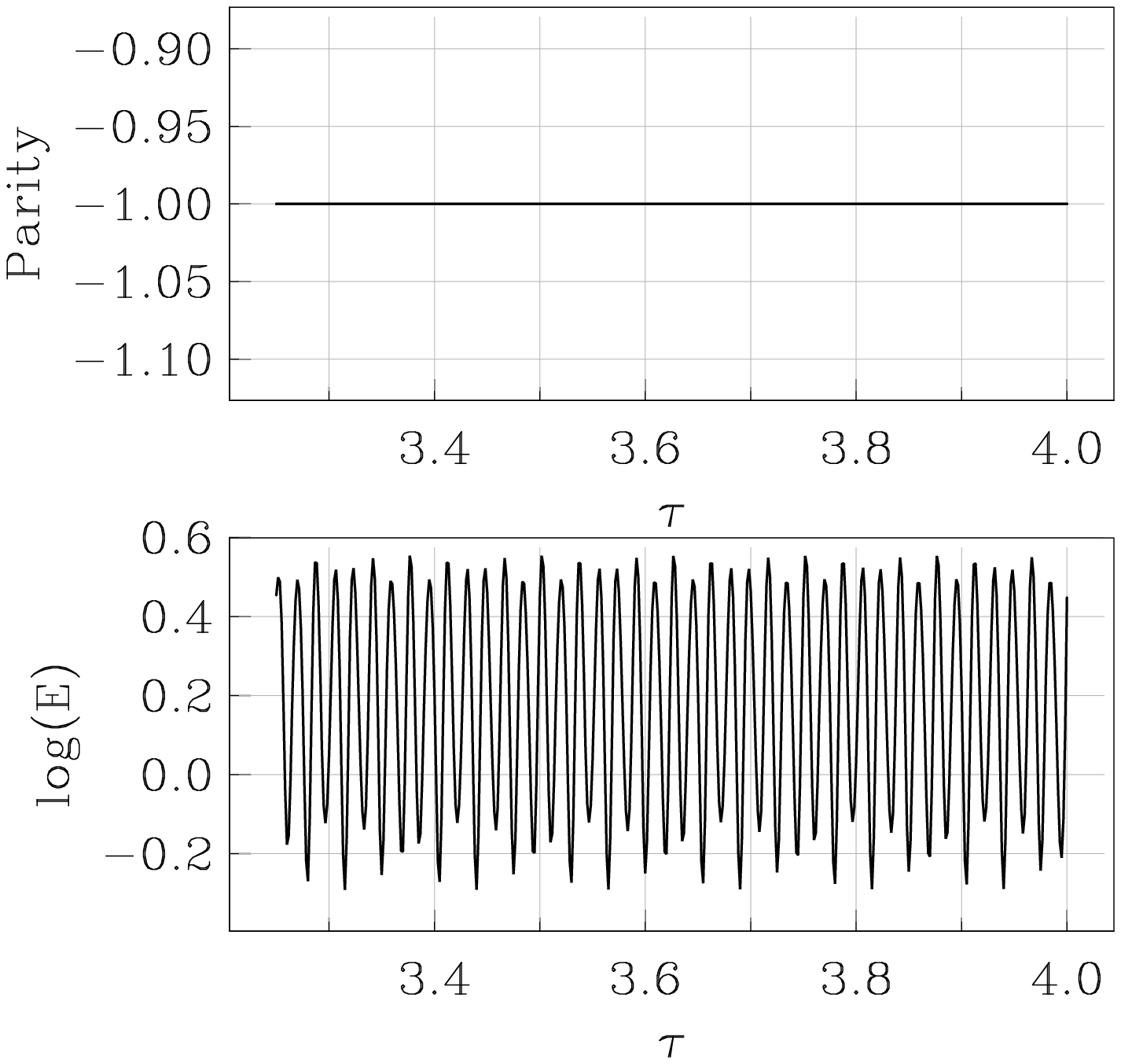} 
\includegraphics[width=0.45\linewidth]{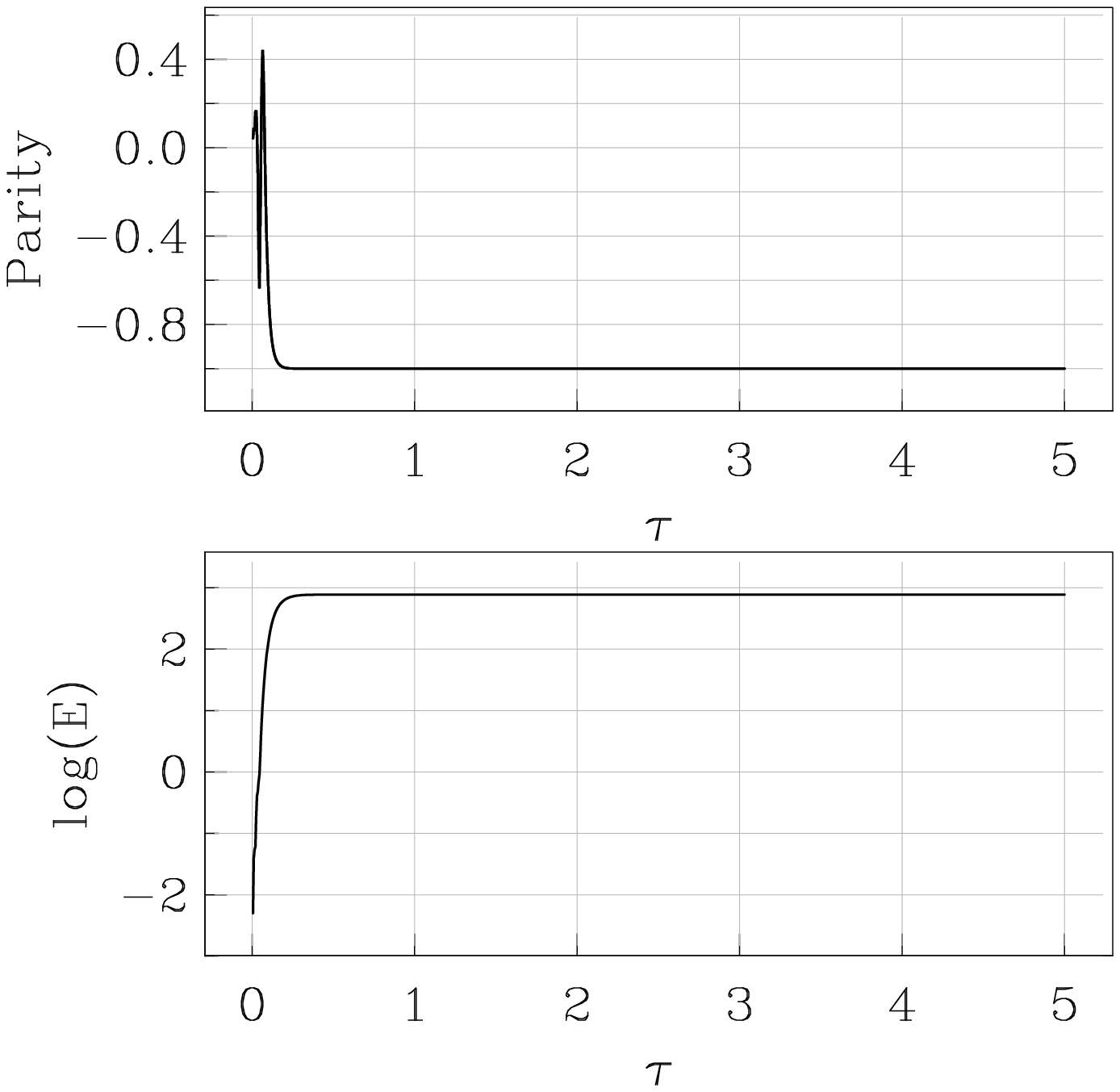} b)\\[0.1cm] 
\caption{Results of dynamo modelling with the rotation law of Fig.~1a.
a) with solar-like differential rotation, $D<0$, \\
b) with anti-solar type differential rotation, $D>0$.
A parity of $+1$ corresponds to a quadrupole-like magnetic geometry,
$-1$ to dipole-like. $E$ is a measure of the global magnetic energy
and $\tau$ is dimensionless time.}
\label{dynmod}
\end{figure*}

We made several experiments in the spirit of \cite{Kea18}, to test the
applicability of these ideas to giant  stars with deep subsurface convection zones, such as HK Lac. We began with a modification of the synthetic solar
rotation law used by \cite{Jea08}, now extended throughout the deeper
convection zone with base at fractional radius 0.6 and nominal overshoot layer thickness 0.05, see Fig.~1a.
If we take the absolute values of the dynamo number used in the solar case by 
\cite{Kea18}, then both negative and positive values of $D$
give oscillatory solutions. If we take the hint that the dynamo number of
HK Lac  may be  much more supercritical, and so double the absolute value of the
dynamo number, then with $D>0$ the dynamo is steady, and with $D<0$ it is
oscillatory (see Fig.~3).
If we now take the rotation law of Fig.~1b, we find a very similar result --  
see Fig. 2.
From these experiments we can deduce that, as in the solar case examined by \cite{Kea18}, we can find dynamo models in which anti-solar differential rotation
gives steady solutions of significantly higher 
magnetic energy than found in the standard oscillatory solutions.
Results are sensitive to some details of the rotation law, and to the value of the dynamo number.

\section{Large flares on some other late-type stars}

As far as we know, the large $H_\alpha$ flare on HK Lac is the only example of a possible superflare (estimated total released energy of approximately $10^{39}\;$erg) reported on a giant star with well-known rotational parameters and decades-long photometric record of its magnetic activity. 

Similar large flares were observed by the {\sl Kepler} satellite on KIC 2852961 \cite{B15}, consisting of 4 or 5 events with total energies between $10^{37}-10^{38}\,$ergs each,  in the short-cadence mode, during 29 days, see Table 1. Several other flares were observed in the long-cadence mode as well, lasting from several hours to about a day.  This star was also observed by the ASAS survey \cite{P02} and those data revealed an average rotational period of 35.58 days. 
Not much is known about this star, but because of its rotational period (35.58 days, ASAS), 
but its $T_{\rm eff} = 4722\;$K, log g = 2.919 (KIC input catalog), and V--I color index 1.15 (ASAS), 
it is very probably a giant.

\begin{figure*} 
\centering
\includegraphics[width=0.75\linewidth]{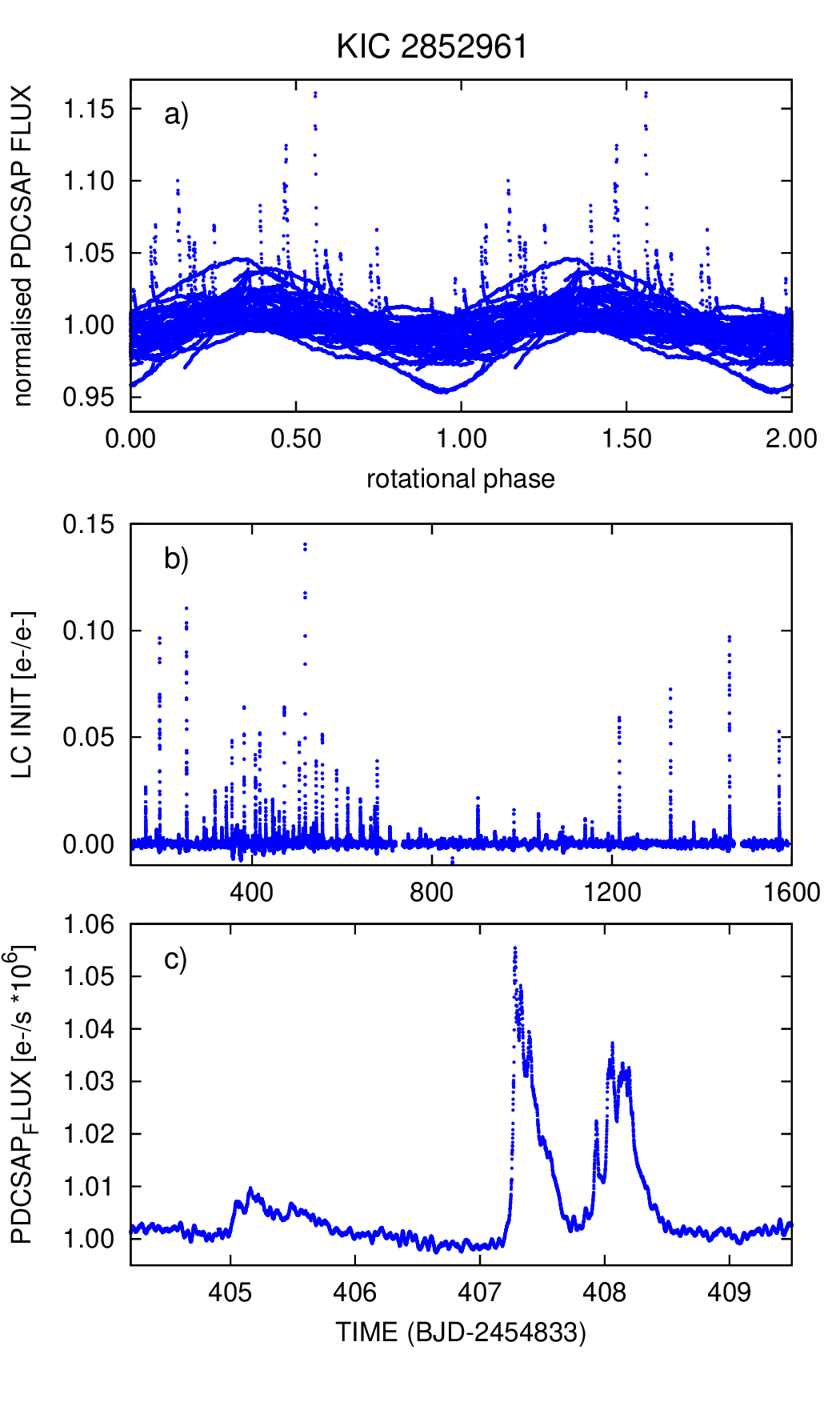}\\[-0.3cm]  
\caption{Superflares on KIC 2852961: \\
a) Normalized Kepler data of KIC 2852961 phased with a rotational period of 35.58 days given by ASAS, \\
b) Detrended, validated Kepler data (DR 25 DV, Q1-17) of KIC 2852961. Note, that the $~36$ days rotational modulation is also subtracted from the dataset, therefore the flare activity is well seen, \\
c) The last three flares from Table 1 \cite{B15}, from the Kepler short-cadence data of KIC 2852961. These flares erupted close to each other in time, probably the last two make up just one, complex flare event.}
\label{KIC}
\end{figure*}

\begin{table*}
\large\centering
\begin{tabular}{|c|c|c|c|c|c|} 
\hline
 KIC  &         BJD    &    log(EW) & log(DF) & Dt   &    log(E) \strut\cr
  
  \hline
              
 2852961 &  5222.707 &  --1.308 &  --2.301 &  14.892 &  37.207 \cr
 2852961 &  5234.205 &  --1.584 &  --2.253 &  \phantom{0}6.980 &  36.930 \cr
 2852961 &  5238.159 &  --0.750 &  --2.105 &  25.109 &  37.765 \cr
 2852961  & 5240.281 &  --0.336 &  --1.265 &  14.058 &  38.179 \cr
 2852961  & 5241.062 &  --0.276 &  --1.414 &  19.371 &  38.239 \cr
\hline 
\end{tabular}
\vspace{0.3cm}
\caption[]{Superflares on KIC 2852961 \cite{B15}: KIC -- Kepler Input Catalog identifier, BJD -- Julian date of peak of flare,  log(EW) -- the total integrated flux of the flare (ergs),   
log(DF) -- the relative flare intensity,  Dt --  the total flare duration (hrs),   log(E) -- the total flare energy (ergs)}
 
\end{table*}

A long-duration (9 days) optical flare with $E\approx 1.8 \times 10^{39}\;$erg  is reported on the fast 
rotating ($P_{\rm rot}= 9.55$ days) FK Com type star YY Men = HD 32918 (K1 III) by \cite{Cea92}.

For the primary of the close binary  II Peg = HD 224085 (K2 IV,
$P_{\rm orb} = 6.7$ days) a hard X-ray flare with $E\approx 10^{38}$ erg (Swift/BAT data) is reported 
\cite{Oea07};  in the cited paper superflares in radio 
wavelengths on two other subgiant stars  (HR 1099, UX Ari) are also mentioned. All three systems belong to 
the  RS CVn type.

For  CF Tuc = HR 5303 (G0 IV+K4 IV),  the longest (lasting 9 days, also with an exceptio\-nal 
rise time of 1.5 days) coherent stellar X-ray flare ever observed by ROSAT has $E\approx 1.4 \times 10^{37}$ 
erg \cite{KS96}. This is also a RS CVn-type partially eclipsing binary system \cite{KS96}.  

For the active eclipsing binary system  SZ Psc (F8 V+K1 IV) of RS CVn type \cite{Xetal16}
a  flare with $E\approx 4.5 \times 10^{36}$ erg is observed by IUE  in ultraviolet  \cite{Dea94}. For this 
system, there are observations of strong flares in optical and X-ray ranges as well.

In contrast to the HK Lac case, we have no observational evidence to support either
solar or anti-solar rotation laws for these stars. The general impression is that large flares on giants and subgiants are not frequent events. However the flare observed on HK Lac is not an isolated example. This seems consistent with a dynamo explanation of the phenomenon as presented in this paper.

\section{Discussion and conclusions}

Our dynamo model demonstrates that a conventional dynamo based on differen\-tial rotation and mirror asymmetry of stellar convection can 
produce sufficient magnetic energy to support such superflares, provided that the differen\-tial rotation is anti-solar. Because the differential rotation of HK Lac 
seems to be anti-solar, dynamo modelling opens the  possibility of producing 
sufficent magnetic energy on HK Lac (and similar active giants in close binaries) to get such superflares.
Our models are deficient in that they do not produce unsteady behaviour
-- however we note that observational data for HK Lac is  rather uncertain.

Of course, we cannot {\it rigorously} claim that the actual interior rotation law of HK Lac is such as to support
the dynamo state discussed above, simply because there are 
no detailed observational data concerning the differential rotation of HK Lac, and the theory of differential 
rotation of binaries and giants is not sufficiently developed.  
\cite{Cetal14} discuss numerical modelling of the convection
zone hydrodynamics that can produce anti-solar surface differential rotation,
but inclusion of their results in our modelling does not seem feasible.
However the results of dynamo modelling suggest this possibility, at least in principle.

We note that we need some (arguably plausible) tuning of parameters to obtain the desired configuration for the dynamo modelling. Our interpretation is that this means that 
superflaring star  such as HK Lac will be rather rare in binaries containing giants. 
This looks consistent with the fact that such superflares are indeed quite rare events in 
previous searches for superflares in binary systems with giant primaries.  On the other hand, this may be 
just a selection of brightness observational effect since giant stars are not monitored continuously because 
of their long rotational periods -- tens of days -- so observing a flare is not very probable. Of course, 
monitoring would be very helpful here as well as in many other aspects of stellar magnetic activity studies.

Any progress in the theory of differential rotation of binaries as well as in theory of stellar flares 
clearly would also be very helpful in understanding 
superflaring giants in binary systems. 
However development of such a theory is obviously far beyond the scope of this paper.
We stress again that we have only attempted to demonstrate the possibility 
certain anti-solar rotation laws are consistent with the production of high 
energy dynamo regimes, and suggest that this is a possible explanation of the
superflaring phenomenon. It is beyond the scope or ambition of this paper
to investigate a suite of rotation laws, and even less so to produce
self-consistent models.

\phantomsection

\section*{Acknowledgments}
\addcontentsline{toc}{section}{Acknowledgments} 

DS acknowledges financial support from RFFI under grant 17-02-00300 and 18-02-00085. 
LLK is thankful to the Russian Foundation for Basic Research for support (project 17-52-80064). 
This research has made use of the NASA Exoplanet Archive, 
which is operated by the California Institute of Technology, 
under contract with the National Aeronautics and Space Administration 
under the Exoplanet Exploration Program.

\phantomsection
\bibliographystyle{unsrt}

\end{document}